\documentclass[english,aps,preprint]{revtex4}
\usepackage[T1]{fontenc}
\usepackage[latin9]{inputenc}
\usepackage{amsmath}
\usepackage{amssymb}
\usepackage{graphicx}

\makeatletter
\@ifundefined{textcolor}{}
{%
 \definecolor{BLACK}{gray}{0}
 \definecolor{WHITE}{gray}{1}
 \definecolor{RED}{rgb}{1,0,0}
 \definecolor{GREEN}{rgb}{0,1,0}
 \definecolor{BLUE}{rgb}{0,0,1}
 \definecolor{CYAN}{cmyk}{1,0,0,0}
 \definecolor{MAGENTA}{cmyk}{0,1,0,0}
 \definecolor{YELLOW}{cmyk}{0,0,1,0}
}

\makeatother

\usepackage{babel}
\begin{document}
\title{Radiation Pressure Induced Oscillations of an Optically Levitating
Mirror}
\author{Satyam Shekhar Jha$^{1,2}$, Tal Carmon$^{3}$,  Fan Cheng$^{3},$Lev
Deych$^{1,2}$.}
\affiliation{$^{1}$Queens College, Department of Physics, Flushing, NY, 11366,
USA.}
\affiliation{$^{2}$Graduate Center of CUNY, 365 5th Ave, New York, NY, USA}
\affiliation{$^{3}$School of Electrical Engineering, Tel Aviv University, Tel
Aviv, Israel}
\begin{abstract}
Optical Fabry-Perot cavity with a movable mirror is a paradigmatic
optomechanical systems. While usually the mirror is supported by a
mechanical spring, it has been shown that it is possible to keep one
of the mirrors in a stable equilibrium purely by optical levitation
without any mechanical support. In this work we expand previous studies
of nonlinear dynamics of such a system by demonstrating a possibility
for mechanical parametric instability and emergence of the ``phonon
laser'' phenomenon. 
\end{abstract}
\maketitle

\section{Introduction}

The idea that cavity optomechanical phenomena can be observed in a
cavity formed by optically levitating mirror has been proposed and
realized by a number of authors\cite{Singh2010,Guccione2013,Ma2020}.
The authors of Ref. \cite{Lecamwasam2020} derived main equations
describing the motion of the center of mass of a levitated mirror
and studied stability of its dynamics. They found that even if the
system possess a stable equlibrium in the quasi-static approximation,
the oscillations around this equilibrium become unstable if one takes
into account corrections to the quasi-stationary approximation. These
corrections result in optically induced amplification of mechanical
motion and run-away instability. In this paper we show that the mechanical
dissipation of the mirror, which is present due to the air resistance
and can be controlled, stabilizes the nonlinear dynamics of the mirror
resulting in a multimode phonon lasing-like behavior\cite{Rivlin1992,Vahala2009,Mahboob,Yamaguchi2014,Okuyama2014,Pettit2019,Sharma2022,Behrle2023,ElSayed2023}.
In this note we present the results of our analysis of the periodic
oscillations of the levitating mirror in this regime. 

Equations of motion for optical and mechanical degrees of freedom
can be written down as

\begin{equation}
M\frac{d^{2}x}{dt^{2}}+\gamma_{m}M\frac{dx}{dt}=-Mg-\hbar\frac{d\omega_{c}}{dx}a^{\dagger}a\label{eq:mech-eq-motion-2}
\end{equation}

\begin{equation}
\frac{da}{dt}+\left(-i\left(\omega_{L}-\omega_{c}(x)\right)+\frac{\pi^{2}}{\mathcal{F}}\frac{c}{x}\right)a=\pi\sqrt{\frac{2c}{\mathcal{F}x}}a_{in},\label{eq:field-annih-op-1}
\end{equation}
where in Eq. \ref{eq:mech-eq-motion-2} $x,M,\gamma_{m}$ are coordinate,
mass, and mechanical damping coefficients of the levitated mirror,
$g$ is the acceleration of gravity, $\omega_{c}$ is the optical
resonance frequency dependent on the mirror coordinate $x$
\begin{equation}
\omega_{c}=\frac{\pi N}{x},\label{eq:cavity-freq}
\end{equation}
($N$ is the order of the optical resonance), and $a$ is the amplitude
of the optical mode. Additionally, parameter $\omega_{L}$in Eq. \ref{eq:field-annih-op-1}
is the frequency of the driving laser, $\mathcal{F}$ is the cavity
finesse, $c$- speed of light, and the term $\pi^{2}c/\mathcal{F}x$
represents the cavity decay rate, while $a_{in}$ describes the amplitude
of the driving field normalized such that 
\begin{equation}
\hbar\omega_{L}\left\langle a_{in}^{\dagger}a_{in}\right\rangle =P_{in}\label{eq:incident-field}
\end{equation}
where $P_{in}$ is the input power. Unlike standard optomechanical
models\cite{Carmon2005,Kippenberg:07}, Eq. \ref{eq:mech-eq-motion-2}
lacks a mechanical spring force, so that the oscillations of the mirror
occur solely due to so-called ``optical spring'' effect\cite{Aspelmeyer2014}.
It shall be noted that while mechanical spring is linear and instantaneous,
the ``optical spring'' is non-linear with respect to mirror's displacement,
and is also characterized by a time delay determined by the lifetime
of the optical cavity mode. 

If the input power $P_{in}$ exceeds the critical value $P_{cr}=Mgc\pi^{2}/(2\mathcal{F})$,
the levitated mirror can be in the stable equilibrium in the position
with coordinate $x_{eq}$ defined as

\begin{equation}
x_{eq}\approx x_{L}+\xi\sqrt{\frac{P_{in}}{P_{cr}}-1}\label{eq:equlibrium}
\end{equation}
where 
\begin{align}
x_{L} & =\frac{Nc\pi}{\omega_{L}}=\frac{N\lambda_{L}}{2}\label{eq:xL}
\end{align}
and parameter 
\begin{equation}
\xi=\frac{c\pi^{2}}{\mathcal{F}\omega_{L}}=\frac{\pi\lambda_{L}}{2\mathcal{F}}\label{ksi}
\end{equation}
characterizes the linewidth of the cavity resonance expressed in terms
of the wavelength of the laser $\lambda_{L}$ rather than inn terms
of the frequency. The corresponding equilibrium field amplitude is
given by
\begin{equation}
a_{eq}=\pi\sqrt{2\frac{c}{\mathcal{F}x}}\frac{a_{in}}{-i\left(\omega_{L}-\frac{Nc\pi}{x_{eq}}\right)+\frac{\pi^{2}c}{\mathcal{F}x_{eq}}}\label{eq:aeq}
\end{equation}

It is convenient to rewrite the equations of motion in terms of real
and imaginary parts of the relative deviation of the field amplitude
from its equilibrium value $a_{eq}$: $w=Re\left[\left(a-a_{eq}\right)/a_{eq}\right]$
and $y=Im\left[\left(a-a_{eq}\right)/a_{eq}\right]$, and dimensionless
mechanical displacement expressed in terms of the cavity line width
$\xi$, $u=\left(x-x_{eq}\right)/\xi$ :
\begin{eqnarray}
\frac{d^{2}u}{d\tau^{2}}+2\eta\frac{du}{d\tau}-2w & = & w^{2}+y^{2}\label{eq:u}\\
\epsilon\frac{dw}{d\tau}+w+ry & = & -uy\label{eq:w}\\
\epsilon\frac{dy}{d\tau}-u+y-rw & = & wu\label{eq:y}
\end{eqnarray}
Here $\tau$ is dimensionless time defined as $\tau=t\varOmega_{max},$where
\begin{equation}
\varOmega_{max}=\sqrt{\frac{g}{\xi}}\label{eq:variables}
\end{equation}
is the maximum mechanical frequency of linear oscillations of the
mirror, $2\eta=\gamma_{m}/\varOmega_{max}$ is the dimensionless mechanical
damping parameter, and $r$ is dimensionless detuning from the equlibrium
position of the mirror $x_{eq}$ defined as 
\[
r=\frac{x_{eq}-x_{L}}{\xi}=\sqrt{\frac{P_{in}}{P_{cr}}-1.}
\]
Parameter $r$ also serves as a measure of input power and is the
main parameter controlling the behavior of the system.  Eq. \ref{eq:u}
- \ref{eq:y} are similar to equations derived in Ref. \cite{Lecamwasam2020}
with one significant difference: Eq. \ref{eq:u} - \ref{eq:y} contain
the term proportional to mirror's velocity $du/d\tau$, which is responsible
for mechanical damping. In the absence of this term the authors of
Ref. \cite{Lecamwasam2020} correctly predicted that optomechanical
interaction results in amplification and the loss of stability of
mechanical oscillations. However, as we will show below, the mechanical
damping stabilizes nonlinear dynamics of the mirror resulting in a
behavior similar to phonon lasing \cite{Behrle2023,Pettit2019,Sharma2022,Parsa2023,Yamaguchi2014,Xiao2020}. 

\section{Nonlinear dynamics of the mirror in the presence of mechanical dissipation}

\begin{figure}

\includegraphics[scale=0.6]{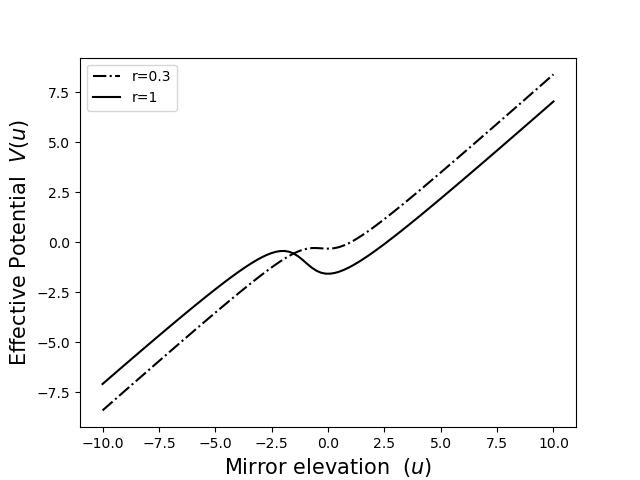}\caption{\label{fig:Optomechanical-potential}Optomechanical potential for
different values of the dimensionless detuning (input power parameter)
$r$. Obviously larger $r$ results in the deeper potential well. }

\end{figure}

The cavity dynamics is controlled by parameter $\epsilon$: $\epsilon=Q_{c}\left(\omega_{max}/\omega_{L}\right)$,
where $Q_{c}=x_{eq}/\xi\approx x_{L}/\xi\gg1,$is the cavity's quality
factor. Terms explicitly containing small parameter $Q_{c}^{-1}$
in Eq. \ref{eq:u} - \ref{eq:y} have been neglected. With $\omega_{L}\sim10^{14}Hz$,
$\mathcal{F}\sim10^{4},$and resonance order $N=200,$we can estimate
$\xi\sim10^{-10}m,$ $\omega_{max}\sim10^{5}Hz$ and $Q_{c}\sim10^{6},$which
yields $\epsilon\sim10^{-3}.$ It is justified, therefore, to analyze
the system of Eq. \ref{eq:u} - \ref{eq:y} using a perturbation expansion
in small parameter $\epsilon$. The same approach was used in Ref.
\cite{Lecamwasam2020} and earlier in Ref. \cite{Kippenberg:07}.
The zero-order solution results in the well known quasi-stationary
approximation, where the optical amplitude is assumed to follow the
mechanical displacement. In terms of the variables used in this work
it is written as
\begin{eqnarray*}
y & = & \frac{u}{1+\left(r+u\right)^{2}}\\
w & = & -\frac{u\left(r+u\right)}{1+\left(r+u\right)^{2}}
\end{eqnarray*}
\begin{align}
\frac{d^{2}u}{d\tau^{2}}+2\eta\frac{du}{d\tau}+\frac{ru}{1+\left(r+u\right)^{2}} & =0\label{eq:zero-order}
\end{align}
In the absence of the damping, the mechanical motion in this approximation
can be characterized by potential
\[
U=\frac{g}{x_{L}}\left[x+\frac{P_{in}}{P_{cr}}\xi\arctan\left[\frac{x_{L}-x}{\xi}\right]\right]
\]
shown in the Fig.\ref{fig:Optomechanical-potential} for several values
of the detuning $r$. In the linear approximation we have harmonic
oscillations with frequency 

\begin{figure}
\includegraphics[scale=0.4]{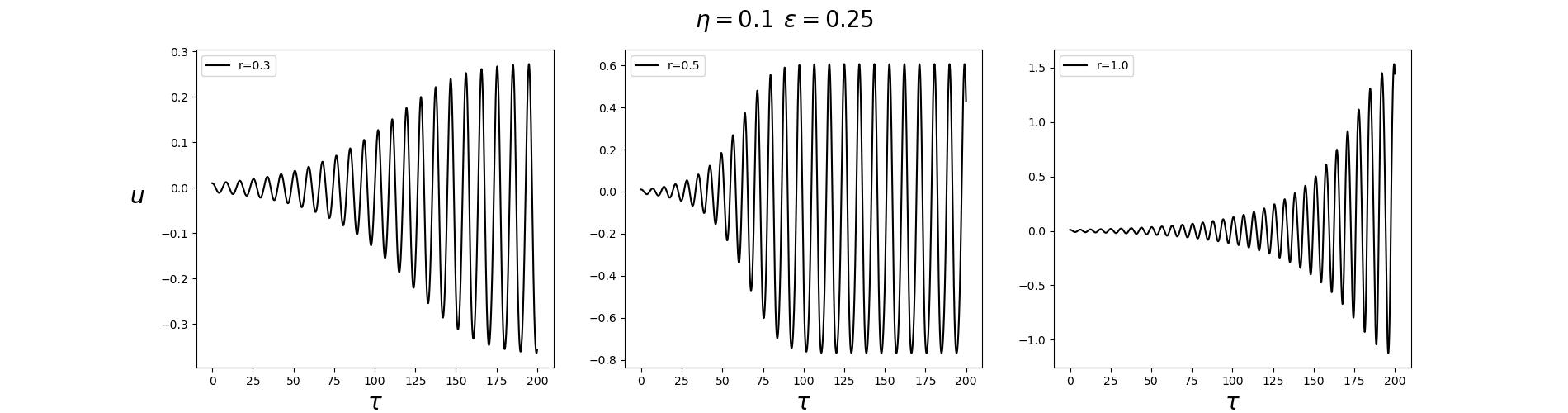}\caption{ Mechanical \label{fig:-Mechanical-displacement}displacement $u$
as a function of time for different value of $r$. The values of mechanical
damping $\eta$ and the cavity dynamical parameter $\varepsilon$
are given in the figures. }
\end{figure}

\begin{gather}
\varOmega_{M}^{2}=2\varOmega_{max}\frac{r}{1+r^{2}}=2\varOmega_{max}\frac{P_{cr}}{P_{in}}\sqrt{\frac{P_{in}}{P_{cr}}-1}.\label{eq:harmonic-freq}
\end{gather}
where the maximum frequency corresponds to $P_{in}=2P_{cr}$. These
oscillations arise solely due to the ``optical spring'' effect as
no mechanical springs are present in the system. The first order correction
in $\epsilon$ introduces an optical amplification, and the nonlinear
dynamics of the mirror in this approximation is described by equation
\begin{equation}
\frac{d^{2}u}{d\tau^{2}}+2\eta_{eff}\frac{du}{d\tau}+\frac{u\left(u+2r\right)}{1+\left(r+u\right)^{2}}=0\label{eq:nonlinear equation}
\end{equation}
where effective dissipation/gain parameter $\eta_{eff}$ is given
by 
\begin{equation}
\eta_{eff}=\eta-2\epsilon\frac{\left(1+r^{2}\right)\left(r+u\right)}{\left[1+\left(r+u\right)^{2}\right]^{3}}\label{eq:nonlinear amplification}
\end{equation}
If $\eta=0$, this parameter is always negative and the optomechanical
interaction results in an unsaturated mechanical gain rendering the
system unstable \cite{Lecamwasam2020}. If, however, $\eta\neq0$,
the strength of the initial linear amplification occuring when the
linear gain parameter 
\begin{equation}
\eta_{eff}^{(lin)}\approx\eta-2\epsilon\frac{r}{\left(1+r^{2}\right)^{2}}.\label{eq:linear-dumping}
\end{equation}
becomes negative is limited by nonlinear terms, and one can expect
an initial growth of the mechanical amplitude to saturate and the
system to settle in stable oscillations. The range of parameters allowing
for amplification to happen is determined by inequality\footnote{We note that for realistic values of the parameters and small enough
$\eta$ Eq. \ref{eq:amplification} is consistent with condition $\epsilon\ll1$ } 
\begin{equation}
\frac{\eta}{2\epsilon}<\frac{r}{\left(1+r^{2}\right)^{2}},\label{eq:amplification}
\end{equation}
The function on the right-hand side of this inequality has a maximum
value $3\sqrt{3}/16$ at $r^{2}=1/3$, so that the amplification regime
is possible only if $\eta/2\epsilon$ does not exceed this value.
For each value of $\eta/2\epsilon$<$3\sqrt{3}/16$, the amplification
region is limited by $r_{min}<r<r_{max},$where $r_{min}$ and $r_{max}$
are solution to equation
\begin{equation}
\frac{\eta}{2\epsilon}=\frac{r}{\left(1+r^{2}\right)^{2}}.\label{eq:r_min-r_max}
\end{equation}
Not surprisingly, for $r$ close to one of the boundaries, when amplification
is weak, the time it takes for system to reach the steady state oscillations
is longer than for $r$ closer to the center of the amplification
region. This point is illustrated in Fig. \ref{fig:-Mechanical-displacement},
 which shows the saturation of the initial amplification as the nonlinear
terms in Eq. \ref{eq:nonlinear equation} stabilize the effective
gain for different values of $r$. One can clearly see the tendency
for the increased time to steady state oscillations for values of
$r$ closer to the boundaries of the amplification region, which for
the chosen values of parameters $\eta/\epsilon=$ is between $r_{min}=0.194$
and $r_{max}=1.298$. 

\begin{figure}
\includegraphics[scale=0.6]{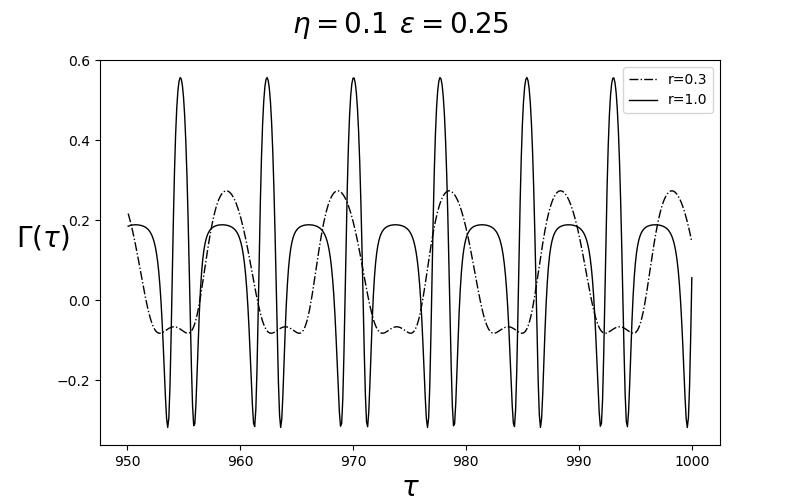}

\caption{Time-dependence of the effective nonlinear gain parameter for two
different values of detuning (power) parameter r. One can see that
the effective gain does not saturate to zero but keeps oscillating
resulting in the energy flowing from light to mechanical degrees of
freedon when the gain is negative and in the opposite direction when
it is positive. \label{fig:Time-dependence-of-the-gain}}
\end{figure}

The observed nonlinear stabilization of oscillations is reminiscent
to the population inversion saturation in regular lasers, but there
is also a significant difference. In a simplest single mode laser
the population inversion saturates to a constant value such that the
effective gain in the steady lasing regime remains zero and the steady
state oscillations are harmonic. In the case considered in this work
the situation is more complicated - the effective gain oscillates
around zero as shown in Fig. \ref{fig:Time-dependence-of-the-gain}.
As a result of these oscillations the steady state regime of the mirror's
oscillations is not monochromatic, and the degree of deviation from
purely harmonic behavior depends on the input power via detuning parameter
$r$. 

\begin{figure}
\includegraphics[scale=0.4]{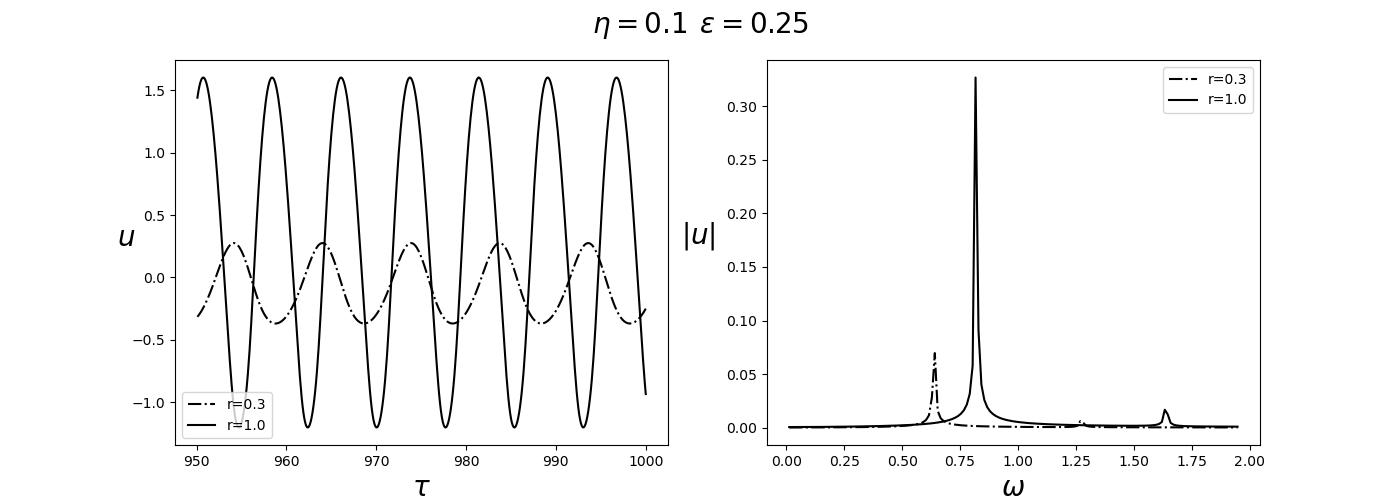}

\caption{\label{fig:Mirror-oscillations-in-steady-state}Mirror oscillations
in the steady state regime for different values of the detuning and
the corresponding Fourier spectra. }
\end{figure}
\begin{figure}
\includegraphics[scale=0.6]{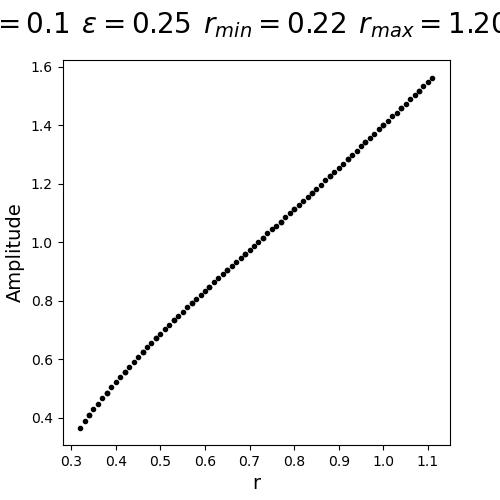}

\caption{\label{fig:Amplitude-versus-detuning}The amplitude versus detuning}

\end{figure}

 This trend is illustrated in Fig. \ref{fig:Mirror-oscillations-in-steady-state}
presenting the time dependence of the mirror's displacement for two
different values of $r$. Oscillations with larger amplitude correspond
to a deeper potential well (see Fig. \ref{fig:Optomechanical-potential}),
which is better approximated by a quadratic behavior, and, therefore,
are more harmonic then the oscillations with smaller amplitude: the
former are characterized by at least three clearly discernible harmonics,
while the latter's spectrum consists of one main frequency with a
weak contribution from another harmonic. Similar trend is also seen
in oscillations of the effective gain parameter shown in Fig.\ref{fig:Time-dependence-of-the-gain}.
The increase in the depth of the potential well with increasing $r$
also explains the corresponding increase in the amplitude of the oscillations
as seen in Figure \ref{fig:Amplitude-versus-detuning}. 

\begin{figure}
\includegraphics[scale=0.4]{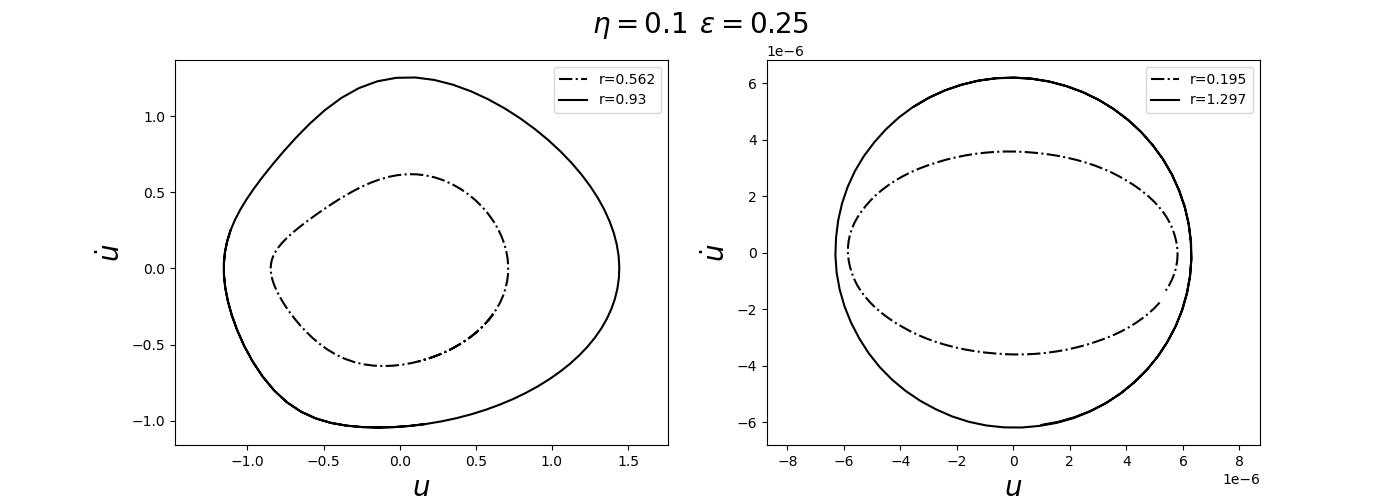}

\caption{\label{fig:Phase-trajectories-of}Phase trajectories of the oscillating
mirror in the steady state regime for strong (left) and weak (right)
nonlinear regimes. }

\end{figure}

\begin{figure}
\includegraphics[scale=0.3]{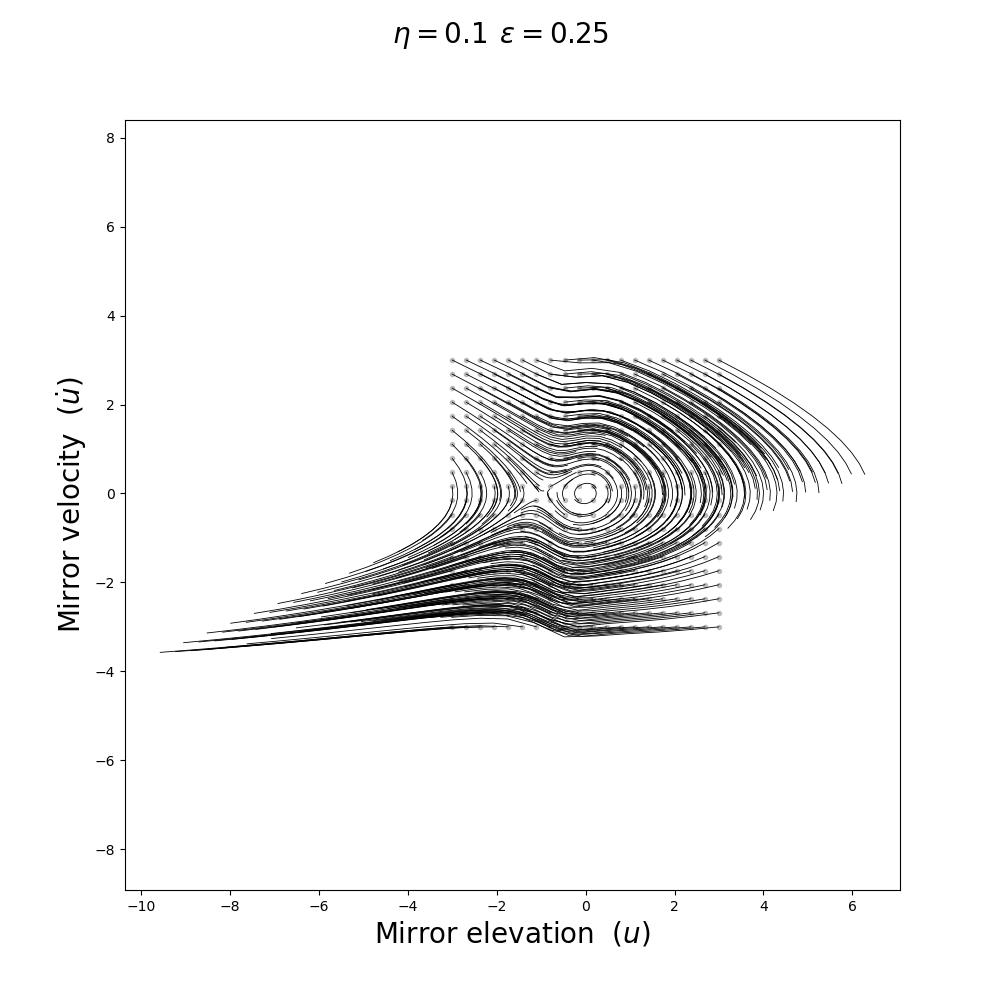}\includegraphics[scale=0.3]{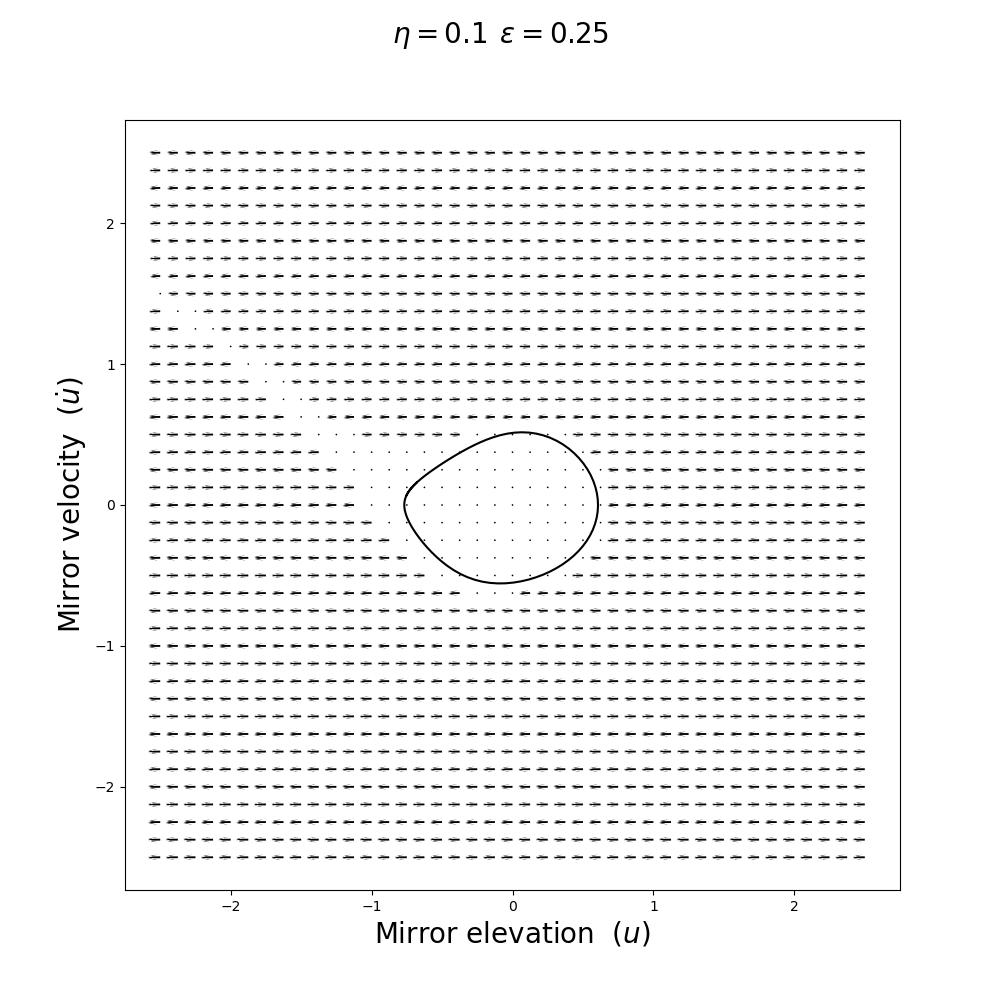}

\caption{(Left) The phase portrait of mechanical oscillations. The stable oscillations
seen in this figure as limiting cycle, approached by the phase trajectories
\label{fig:The-phase-portrait} (Right) The separation of the phase
space into regions of stable limiting cycles, and unstable regions. }

\end{figure}

To further illustrate the nature of the nonlinear oscillations of
the mirror we have constructed the phase trajectories of the oscillations
by plotting velocity $\dot{du/d\tau}$ versus displacement $u$, shown
in Figure \ref{fig:Phase-trajectories-of}. The circular and elliptic
phase trajectories in the left figure are indicative of weakly anharmonic
oscillations, while more complicated shapes in the right panel corresponds
to the stronger anharmonicity. 
 Finally, we present the phase portraits of our oscillating mirror
presenting multiple phase trajectories (Fig. \ref{fig:The-phase-portrait},
left). One can clearly see two types of phase trajectories: those
forming limiting cycles at the center of the figure and corresponding
to stable oscillations, and those that move the system away from the
limiting cycle corresponding to unstable motion of the mirror. This
figure is complimented by a plot on the right showing separation of
the phase space in two regions. The region marked by dots corresponds
to initial conditions resulting in limiting cycle type oscillations,
and those marked by the arrows represent initial conditions resulting
in unstable motion of the mirror. A remarkable feature revealed by
this plot is that there exist initial conditions placing the mirror
outside of the potential well (Fig. \ref{fig:Optomechanical-potential})
but still resulting in stable oscillations in the steady state regime
(points outside of the central region marked by a closed phase trajectory).

\section{Conclusion}

In this work we analyzed the motion of a center of mass of a levitated
mirror in the presence of mechanical damping. We showed that the damping
stabilizes nonlinear dynamics of mirror resulting in a behavior reminiscent
to phonon lasing. Unlike the simplest lasing dynamics, however, the
time dependence of the mirror's displacement in the steady state is
in our case is anharmonic with the effective gain parameter oscillating
around rather then being pinned to zero. 
\begin{acknowledgments}
This research was supported by a NSF -BSF (United States--Israel
Binational Science Foundation ), grant \#2020683 and by Israeli Science
Foundation, grant \# 537/20.
\end{acknowledgments}

\end{document}